# Using ontology for resume annotation

Wahiba Ben Abdessalem Karaa*
High Institute of Management,
41, Rue de la Liberté, Cité Bouchoucha
2000 Le Bardo, Tunis, TUNISIA.
E-mail: wahiba.abdessalem@isg.rnu.tn
*Corresponding author

Nouha Mhimdi
High Institute of Management,
41, Rue de la Liberté, Cité Bouchoucha
2000 Le Bardo, Tunis, TUNISIA.
E-mail: nouhamhimdi@yahoo.fr

*Abstract*—Employers collect a large number of resumes from job portals, or from the company's own website. These documents are used for an automated selection of candidates satisfying the requirements and therefore reducing recruitment costs. Various approaches for process documents have already been developed for recruitment. In this paper we present an approach based on semantic annotation of resumes for e-recruitment process. The most important task consists on modelling the semantic content of these documents using ontology. The ontology is built taking into account the most significant components of resumes inspired from the structure of EUROPASS CV. This ontology is thereafter used to annotate automatically the resumes.

*IndexTerms*—Annotation, e-recruitment, ontology, Semantic web, text analysis, Resume, CV.

## 1 Introduction

Employers often receive a large number of Resumes for an open position. The costs of classically and manually selecting on appropriate candidates have amplified and employers are searching for tools to automate the candidate's selection. Online recruitment processes can be efficient using Semantic Web technologies (Bizer et al., 2005). Indeed, using Semantic Web Technologies in the recruitment domain could considerably decrease the costs for employers: publishing job offers, and selecting candidates (Morin and Trichet, 2004).

In this paper we investigate how Semantic and Web technologies can be used to support the recruitment process. In particular, we describe the design and implementation of an ontology which defines the semantic content of resumes. We propose in addition an annotation approach of resumes based on this ontology. We also present the results of initial experiments testing the performance of this prototype implementation.

The rest of the paper is organized as follows: Section 2 gives a brief overview of related works of the e-recruitment field. We describe our approach for the ontology creation and the global architecture of the annotation system in section 3. In section 4 an evaluation of our approach is shown, and then a conclusion and perspectives for improving this work are given at the end of this paper.

## 2. Related Works

In the field of e-recruitment, the information contained in the resume of a candidate who applies for a job, is specific, and requires semantic and automatic processing. In this context, several approaches conducted under the recruitment (e-recruitment), take into account the semantics of used documents (e.g. resume). Some of them index resumes using semantic techniques that consist mainly on the association of most important elements in a document to concepts existing in the resume. This approach is based mainly on the techniques of Natural Language Processing (NLP).

Other works focus on the semantic annotation approach to enrich documents with their semantic content. This approach seems particularly interesting for its efficiency, and availability of tools and web semantic standards.

*Approach of annotation resumes using Gate*

In order to facilitate the process of e-recruitment, Amdouni and Ben Abdessalem (Amdouni and Ben Abdessalem Karaa 2010), developed a prototype allowing an automatic annotation of the resume. A non structured resume existing in different formats (Doc, Pdf, HTML, etc) can be transformed into an XML document, structured according to the EUROPASS CV constituents. After that, the structured resume is analyzed to extract the pertinent information as the diploma, experience, name, sex, date of birth, nationality, etc. Therefore, they use the API Gate [http://gate.ac.uk/], extended by new JAPE rules permitting the extraction of the necessary information to structure them according to the EUROPASS CV.

The core component of the prototype is the recognition of the basic elements in the resume. For this purpose, they use Gate components: Sentence Splitter, Tokenizer, Gazetteer, and Named Entity Transducer. First, sentence Splitter component identifies sentences in the resume. A Tokenizer splits each sentence into words (punctuation, numbers …). Gazetteer component creates annotations about entities such as persons, organizations, job titles. Named Entity Transducer is used in the Semantic annotation Step. It applies JAPE rules to the previous annotations to generate additional and specific annotations. However, JAPE rules were extended to extract all the key information from resume. Finally an XML document is generated related to HR-XML EUROPASS CV schema.

*Approach of annotation resumes using ontology*

Ontology is gaining position in computer science. It facilitates the understanding between persons and organizations. Ontology is a model, or a theory of a particular domain representing a real-world. Recently, the research on ontologies proves it to be an important field in the Information Systems area (Al-Debei, and Fitzgerald, 2009).

The appliance of ontologies in recruitment domain is relevant, since ontology for recruitment can define concepts such as "competency", "skill", attributes of such concepts, and the relationships between these concepts. In consequence, we can define which competencies and skills are required for a certain job and which experience and knowledge is required to attain a certain level of competence.

Competency is the concept that describes the skills and knowledge that individuals should have to be fit for particular jobs (De Leenheer et al., 2010).

A model which could identify competency is highly needed. A competency model is a set of competency elements in relation and characterising the job performance (Chung and Wu, 2011). A competency modelling is taking considerable interest in different area such as recruitment (Mochol et al., 2006), engineering knowledge (Sicilia et al., 2009), learning (Monceaux et al., 2008), and education (Sgouropoulou, 2010).

The ontology based competency modelling approach (Sicilia, 2005), (Draganidis at al., 2006), describes the desirable competence profile of an employer for a given post. It gives a good definition of the relations between competencies. It can be in addition used for competence assessment process.

Several schemas were defined to describe the competences. HRXML specifies the competency format through XML code (http://www.hr-xml.org/). HRXML enclosing human-resource management data specifications, it has an important use for human resource applications.

The IMS consortium (http://www.imsglobal.org) as well as HRXML offers specifications for competencies named "Reusable Definition of Competency or Educational Objective (RDCEO)".

Ontology offers additional modelling elements, through logical descriptions, for defining in competency schemas different types of competency, different relations, and various measurement scales.

Ontology description languages can be useful for representing both the concepts and the Instances of the concepts.

In the context of recruitment applications, ontologies can be directly used to make applications more aware of the domain semantics, such us matching of an individual's competency profile with a requirements profile, e.g. for applicant selection (Mochol et al., 2006).

*HR ontology*

Bizer et al. (Bizer et al. 2005) have created human resource ontology (HR ontology) composed of sub-ontologies which are used in both job posting and job application descriptions. The HR ontology is derived from existing specifications and standards such as HR-XML standard, HR-BA-XML (German version of HR-XML).

*COmmOnCV project*

Harzallah et al. (Harzallah et al, 2002), worked on the project CommOnCv[1] . This project offers the opportunity for a job seeker or a recruiter to identify and represent the underlying skills in resume or job offer in a formal way. This formal representation in the form of annotations is used as a reference to refine the process of matching between the resume and the adverts from the job market on the web.

The objective of CommOnCV consists on the definition of a model of competency and a skill management process underlying in a resume or a job offer. The competency model focuses on the recognition of skills including social, organizational, technical, and economic aspects. The competencies are characterized by a set of resume's annotations. These latter, are formally denoted by using semantic web languages such as RDF/RDFs or DAML+OIL. These annotations, which follow a model of competency, are described in relation to domain ontologies. These ontologies can be associated to a particular sector: Finance, Healthcare and/or to a specific company.

*OS-SKILL project*

According to Rieu et al (Rieu et al., 2005), OS-SKILL project is a method for managing jobs and skills; it is an Internet / Intranet application that can manage a successful business operator in the repositories of trades and skills. It is fully configurable and allows mapping the trades, skills, organizational structure, career management, the tasks of the organization, etc.

---
[1] http://www.francky-trichet.com/commoncv/English/index.html

OS-SKILL is also a method and a tool: a method of construction trades and skill benchmarks, and a tool for management of jobs and skills.

It provides relevant and effective functionalities independent of management skills, such as staff evaluation, job identification, identification of individuals who need training, analysis situation of the organization, and search for collaborators.

OS-SKILL is based on XML technologies and Web services; it uses the Osia method based on ISO 704, which builds ontology step by step.

*ER-ontology*

In the context of semantic web, Yahiaoui (Yahiaoui et al., 2006) have proposed an approach of semantic annotation of resumes and job offers in order to automate the process of e-recruitment. This is a simple, detailed, and sufficient solution that determines the necessary elements to the process of annotation.

The objective of this work is to offer coherent and clear documents (resume and job offer) codified in XML, the standard for exchanging data on the web.

Ontology is constructed and implemented in the field of computing and telecommunications to model the semantic content of documents and skills underlying documents. The system proposed in this work focuses on an ontology of Human Resource Management (HRM). It contains several sub-ontologies inter-related and generates annotations from their instantiation. It also operates an XML / HTML server to allow storage and management of resumes and job offers. The software interface allows the automatic annotation of documents. It offers the possibility to job seekers to find the offer that fits with their skills.

The user interface for matching allows interpreting the user queries and the calculation of the degree of semantic matching between user documents and annotated documents.

*Job portals*

Mochol et al. (Mochol et al., 2007) construct a prototype job portal using semantically annotated job offers and applicants. They outline how the technique of query approximation can be the basis for a solution of problems in job search.

*Objective work*

In the context of e-recruitment problematic and in view of the fact that ontology is an important component of several applications, we propose an approach for resume annotation founded on resume ontology.

The idea is to build and implement an ontology that regroups basic information in the resume. This ontology will be used, after that, in a semantic annotation process of resumes. The ontology to construct has to follow a specific grammar: HR-XML Schema related to EUROPASS CV.

The semantic annotation is a specific metadata (Sicilia, 2006) generation about resumes, aiming to add them formal descriptions (We can identify attributes describing competencies, occupations, etc.). The semantic annotation of resumes can be useful for many human resource applications: retrieval, classification, extraction of element dependencies, analysis of relationships between elements, etc.

**3. The approach**

This section describes the resume ontology construction steps in Section A. The ontology is baptized **ERECO (E-RECruitment Ontoloy)**. In the Section B is described the automatic resume annotation system, called **E-RecSys** (E-Recruitment System), using ERECO ontology.

*A. Construction of the resume ontology ERECO*

To construct our ontology, we make use of two international standards: HR-XML and EUROPASSCV.

*EUROPASS CV*

The EUROPASS Curriculum Vitae (EUROPASS CV)[2] enables users to document the development of their qualification profile in a standardised format systematically and chronologically. It encloses personal information in addition to details of any education and training, work experience, skills and competences that the individual has. It aims to offer to users a standardized document at European level. It can be filled in online, in all European languages, and in different formats ((PDF, XML, HTML, …). The EUROPASS CV can be considered as an important tool for better transparency of qualifications and competences both for applicants and potential employers at home and abroad. It contains information about individual:

- Personal information: contains eight fields: first name(s)/surname(s), address (es), telephone(s), faxe(s), email, nationality, date of birth and gender.
- Work experience: concerns the occupation or position held, date, and main activities and responsibilities.
- Education and training: describes study contents, date, skills covered, and level in national or international classification.
- Personal skills and competences: this part is dedicated to skills and competences acquired in the individual life. It concerns languages, technical skills and competences, social, organisational, technical skills and competences, computer skills and competences, artistic skills and competences, driving licence, and other skills and competences
- Additional information: comprises any other information that may be relevant, for example contact persons, references, etc.
- Annexes: inventories any items attached.

*HR-XML Schema :*

HR-XML is a set of XML specifications aimed at facilitate the exchange and automated processing of information related to human resources management. The published specifications are free for access.

---

[2] http://europass.cedefop.europa.eu/europass/home/vernav/Europass+Documents/Europass+CV.csp?loc=en_GB.

The standardization of resume is supported by HR-XML. Indeed with HR-XML, a resume may contain the competencies of the applicant in a standardized manner.

In this work, we created a HR-XML schema for resume regarding the EUROPASS CV. Hence, the different items of EUROPASS CV (work experience, education and training, personal skills and competences, etc.) will be described by standard specifications in HR-XML.

Our ontology required to respect the HR-XML schema related to EUROPASSCV.

*The ontology ERECO*

A multitude of ontological engineering methods exists, describing the stages to follow for the construction of ontology. The design of our ontology is guided by the method developed by the University of STANFORD (Noy and Mcguinness, 2002). It includes the following seven steps:

- *Determine the domain and scope of the ontolog.*
- *Consider reusing existing ontologies.*
- *Identify the important terms of the ontolog.y*
- *Define the classes and the hierarchy of classes.*
- *Define the properties of the classes (the attributes).*
- *Define the facets of the attributes.*
- *Create instances of the classes.*

*Step 1: Determine the domain and scope of the ontology*
In this step, we start by answering a set of basic questions that helps to determine the domain and the range of the ontology:

Question: what is the domain that the ontology is going to cover?

Answer: HRM, e-recruitment, International standards (HR-XML and EUROPASS CV) are also a part of the context.

Question: for what we are going to use the ontology?

Answer: for resume annotation

Question: what kinds of questions will the ontology give answers?

Answer: what degree? , What experience, what competence, etc.

Question: who is going to use and to assure the maintenance of the ontology?

Answer: recruiters, applicants.

*Step 2: Consider reusing existing ontologies*

We will assume that no appropriate ontologies subsist and begin developing the ontology from scratch.

*Step 3: Identify the important terms of the ontology*

During this stage, it is important to identify the domain knowledge by establishing a complete list of the terms in the corresponding domain of interest in non structured form. It is necessary to know the concepts, the relations, the rules and the constraints.

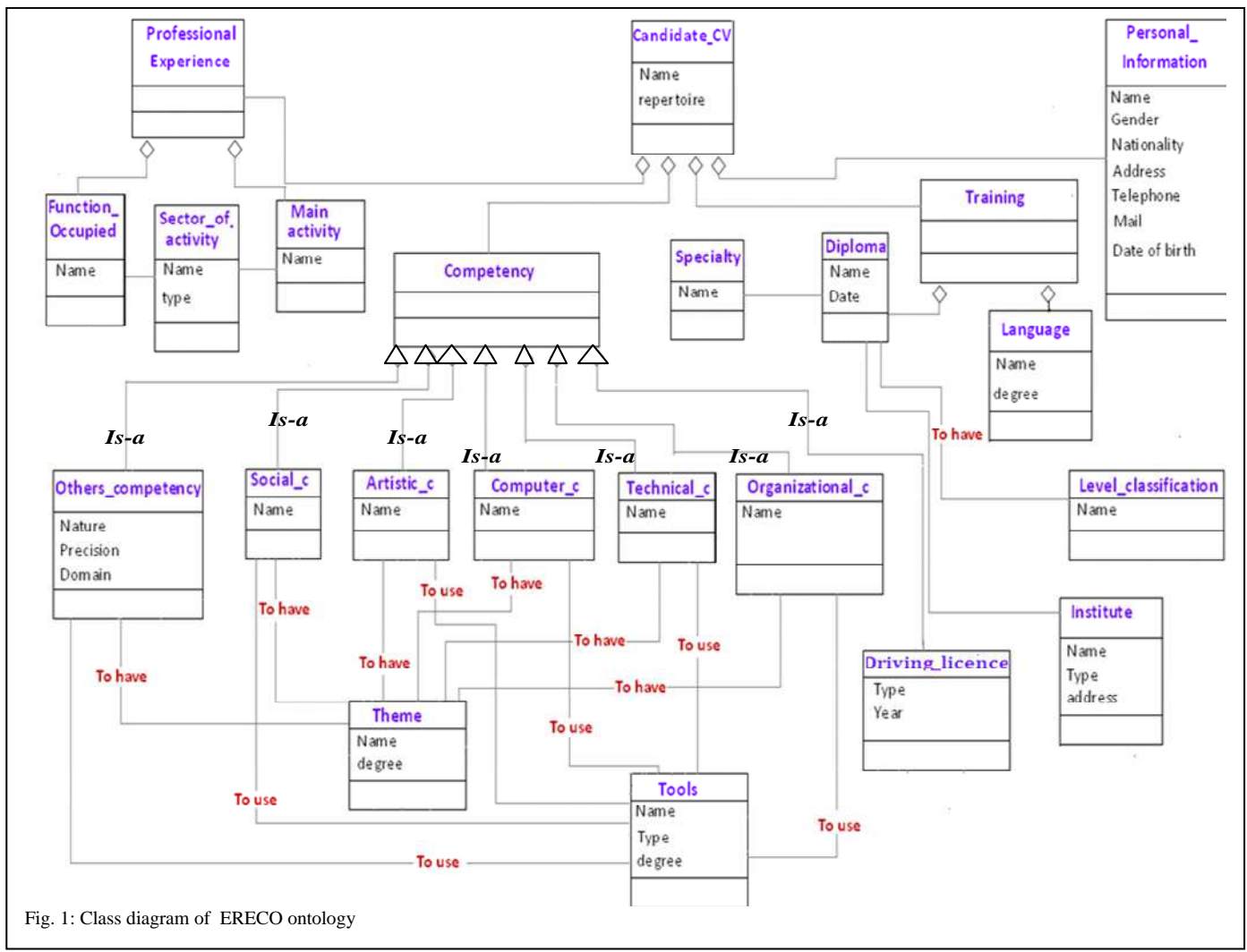

Fig. 1: Class diagram of ERECO ontology

For the resume ontology, we define for example: Profile, age, telephone, professional experience, sex, nationality, education, formation, establishment, language, diploma, university, certificate, level, software, mention, expertise, etc. This step is guided by information contained in the EUROPASS CV.

*Step 4: Define the classes and the hierarchy of classes*

In the previous stage, we identified, for the ontology, a list of non structured terms, now, it is necessary to define classes by selecting the terms that describe the objects having an independent existence. It is necessary to organize these classes in a hierarchical taxonomy.

*Step 5: Define the properties of the classes (the attributes)*

At this stage, it is coherent to assign some properties for the definite classes. The properties will be attached to the classes. If an attribute is connected to the most general class, then all under classes inherit this attribute. Also, it is recommended to affect some values by default to the class attributes.

*Step 6: Define the facets of the attributes*

In this step, we define the value type of the attributes: the types of value capable to be affected to an attribute such as: number, string, character, Boolean…

Also, we designate the number of values (cardinality) that an attribute can have, this cardinality can be unique or multiple. We identify the domain, the rank of an attribute and the relations between classes. The following figure (Fig. 1.) is an UML class diagram showing the design of ERECO ontology.

*Step 7: Create instances of the classes*

This stage consists on creating the instances of the classes that represent real entities.

The ontology population guarantees the efficiency of the annotation. We used the lexicographic data base WordNet for the instantiation of some classes of the ontology in a semiautomatic way.

WordNet is a lexical network developed by the Cognitive Science Laboratory (Princeton University) under the direction of George A. Miller and Christiane Fellbaum [http://www.cogsci.princeton.edu/~wn / \2010]; it covers the names, the verbs, the adjectives and the adverbs. About 150 000 words are organized in wholes of synonyms named synsets permitting to regroup the terms denoting a given concept. In this work we collected for example from WordNet, all the appropriate ontology instances of the concepts "language", "diploma", "nationality", etc. The instances of the other concepts are introduced manually.

ERECO ontology is implemented (Fig 2) using the ontology editor Protégé 2000 [http://protege.stanford.edu]

*B. Annotation system E-RecSys*

The ontology **ERECO** is afterwards exploited in the annotation process of resumes. We proposed a system that we called **E-RecSys** ensuring the annotation of resumes using **ERECO** ontology. The overall system architecture is shown in Figure (Fig. 3.), it consists of five phases:

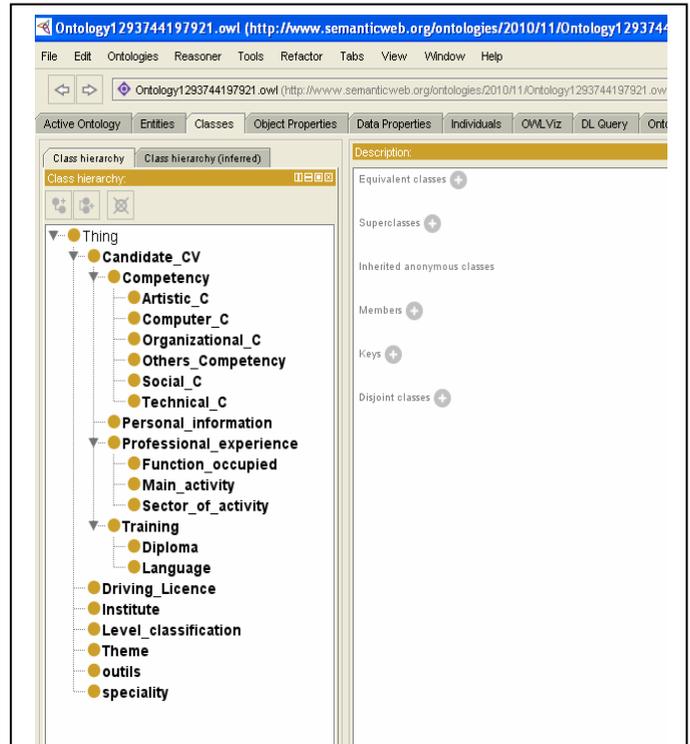

Fig.2.: ERECO Ontology with aptitude and Competencies

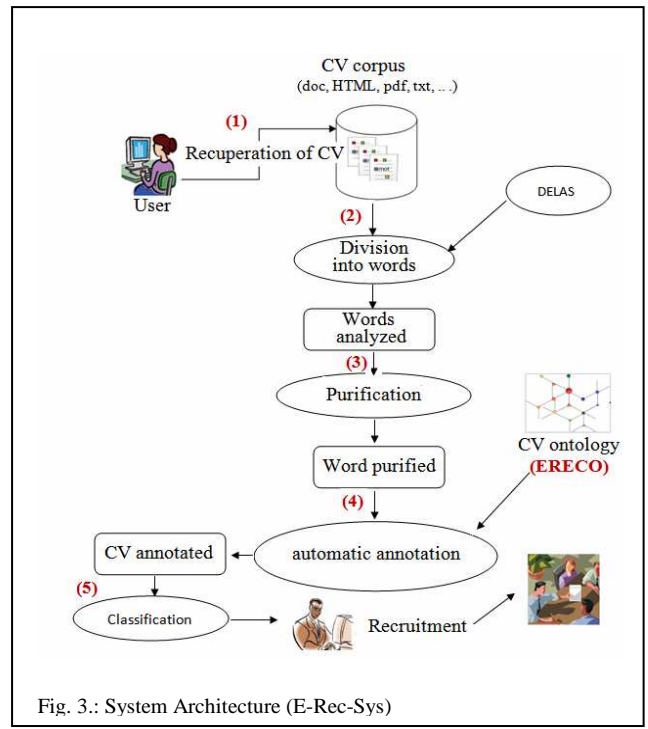

Fig. 3.: System Architecture (E-Rec-Sys)

*Phase 1*

The first phase involves the selection of resumes in a corpus in different formats (doc, XML, PDF, HTML).

*Phase 2*

The second phase is the treatment of resume; it begins by cutting the resume into words, and then analyses each word by identifying its category. In this phase we used a morphological dictionary containing 290,460 entries, which are associated with a morphological code as the coding dictionary DELAS (Courtois and Silberztein, 1990). The morphological dictionary is composed of simple words with no separator, in small letters under their canonical form, indicating the part of speech: verbs (V), names (N), adjectives (A), adverbs (ADV).

*Phase* 3

The third phase refines the list of words, by keeping only the interesting ones (nouns, adjectives, adverbs ...) and omit others like prepositions, and articles.

*Phase* 4

Once the words are refined, in this phase we take advantage of ERECO, which gathers all the concepts that may exist in a resume (experience, competency, training, etc.) and their properties (name, degree, doùain, etc.) and relations, to annotate automatically the resumes.

*Phase* 5

Once the resumes are annotated, we can in this last phase classify them according to different criteria according to the employer's requirements. This classification will be used in finding simply the most suitable candidate.

We opted for java (Jbuilder) as a programming language for the implementation of E-RecSys annotation system. An API (Application Programmation Interface) JAVA allows the access to the ontology within the editor Protégé-2000 (Fig 4).

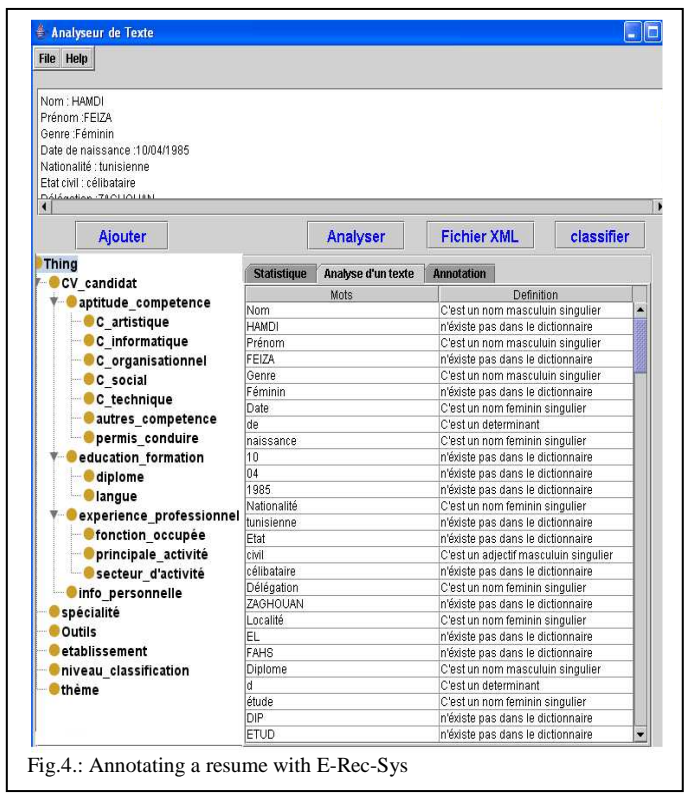

Fig.4.: Annotating a resume with E-Rec-Sys

The result of annotation with E-Rec-Sys can be displayed in an HRXML format (Fig 5).

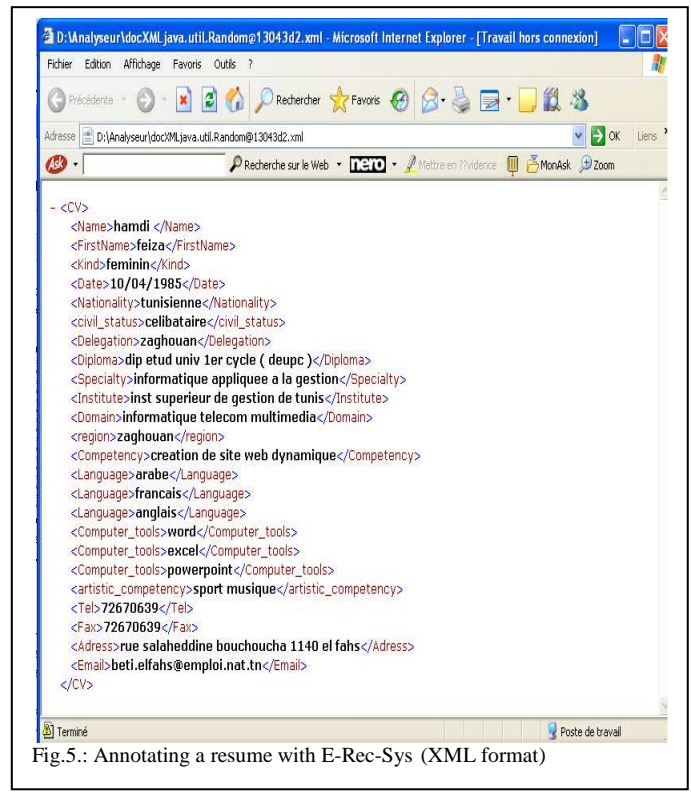

Fig.5.: Annotating a resume with E-Rec-Sys (XML format)

## 4. Evaluation

To evaluate an annotation system, we have to compare the result of the system with the annotation made manually (by an expert)

For the evaluation of E-Rec-Sys, we have annotated the corpus of resumes semi-automatically using the software Gate [http://gate.ac.uk/ ], (Maynard et al., 2000); it is an environment that offers a user the possibility to annotate a document using the plugin Gate Ontology Annotation Tool (OAT). This API allowed us to annotate document (resume) using the ontology ERECO imported with Gate. The concepts of the ontology are displayed, and the expert can associate for each term in the resume the appropriate concept of the ontology (Fig.6).

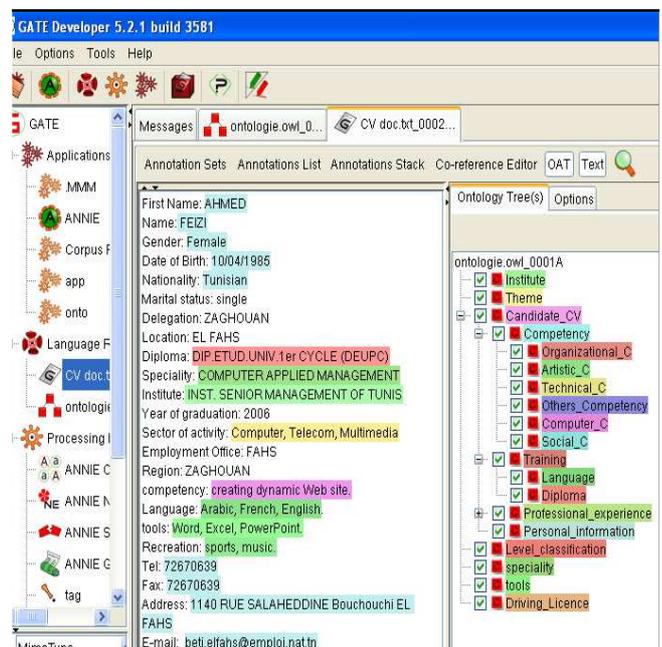

Fig. 6. Annotating using Gate API (OAT)

The two annotated documents (the resume annotated with our annotation system E-Rec-Sys and the resume annotated semi-automatically) are used as inputs for the computation of the statistical metrics: **Precision, Recall, and F-measure.** These parameters are widely used for evaluating such systems. A high recall indicates that you haven't missed anything; a high precision reveals that all returned results were relevant, and F-measure combines exactness (precision) and completeness (recall).

Recall is computed as the fraction of correct annotated words among all truly annotated words:

$$\mathrm{Recall} = \frac{Correct + 1/2\, Partial}{Correct + \mathrm{Missing} + Partial}$$

However, precision is the fraction of correct annotated words among those annotated by the system:

$$precision = \frac{Correct + 1/2\, Partial}{Correct + Spurious + Partial}$$

F-measure combines recall and precision in the following form:

$$F - measure = \frac{(\beta^2 + 1)\mathrm{Precision} * \mathrm{Recall}}{(\beta^2 * \mathrm{Recall}) + \mathrm{Precision}}$$

We used the annotation tool **AnnotationDiff** (Maynard, et al., 2001), which is a Gate plug-in that allows automatic evaluation systems. Indeed, it can compare two sets of annotation; a document annotated automatically with another annotated manually.

The figure (Fig. 7.) shows the result of the evaluation using the document annotated semi-automatically with Gate (CVgate) and the document annotated by E-Rec-Sys (CVxml).

Fig. 7. Annotation Diff result

This result shows that we have six partially correct annotations. The performance metrics Recall, Precision and F-measure are (0, 0, 0) when we consider that the partially correct annotations are false. However, when we consider that the partially correct annotations are correct annotations, the performance metrics Recall, Precision and F-measure turn into (1, 1, 1).

Using the same approach, we complete our evaluation using a corpus gathering 300 resumes downloaded from the site (www.emploi.nat.tn ) and from the site (http://www.e-marketing.fr/Emploi/Recherche/CV-Flash.asp).

The tables below show the performance metrics for personal information (Table I), and other resume information (Table II):

TABLE I
PERFORMANCE METRICS FOR PERSONAL INFORMATION ANNOTATIONS

|           | Name | Gender | Nationality | Date | Adress | Phone | Fax  | Email |
|-----------|------|--------|-------------|------|--------|-------|------|-------|
| Precision | 0.91 | 0.91   | 1           | 0.91 | 0.81   | 0.98  | 0.98 | 0.85  |
| Recall    | 0.81 | 0.87   | 0.97        | 0.95 | 0.85   | 1     | 1    | 0.9   |
| F-mesure  | 0.86 | 0.89   | 0.98        | 0.93 | 0.83   | 0.99  | 0.99 | 0.87  |

TABLE II
PERFORMANCE METRICS FOR OTHER PART ANNOTATIONS

|           | Institute | Language | competency | Training |
|-----------|-----------|----------|------------|----------|
| Precision | 0.8       | 0.97     | 0.78       | 0.81     |
| Recall    | 0.83      | 0.97     | 0.8        | 0.83     |
| F-mesure  | 0.82      | 0.97     | 0.79       | 0.82     |

As shown in tables (Table I) and (Table II), the results are Satisfactory since the values are close to 1. Indeed, the Precision varies between 0.78 and 1; Recall varies from 0.8 to 0.97, and the F-measure between 0.79 and 0.99.

We compared our results with the results of (Amdouni and Ben Abdesslem, 2010) (Table III) and (Table IV). We noted that both works are very close in terms of performance; the difference between metric values is minor.

TABLE III
COMPARISON FROM THE RECALL AND PRECISION

|             | Recall    |                              |             | Precision |                              |
|-------------|-----------|------------------------------|-------------|-----------|------------------------------|
|             | E-Rec-Sys | (Amdouni and Ben Abdesslem, 2010) |        | E-Rec-Sys | (Amdouni and Ben Abdesslem, 2010) |
| Name        | 0.81      | 0.74                         | Name        | 0.91      | 1                            |
| Gender      | 0.87      | 0.94                         | Gender      | 0.91      | 0.83                         |
| Nationality | 0.97      | 1                            | Nationality | 1         | 0.84                         |
| Date        | 0.95      | 0.9                          | Date        | 0.91      | 0.9                          |
| Adress      | 0.85      | 0.86                         | Adress      | 0.81      | 0.81                         |
| Phone       | 1         | 0.97                         | Phone       | 0.98      | 0.97                         |
| Email       | 0.9       | 0.86                         | Email       | 0.85      | 0.91                         |
| Institute   | 0.83      | 0.85                         | Institute   | 0.8       | 0.74                         |
| Language    | 0.97      | 0.88                         | Language    | 0.97      | 0.82                         |
| Competency  | 0.8       | 0.81                         | Competency  | 0.78      | 0.85                         |
| Training    | 0.83      | 0.83                         | Training    | 0.81      | 0.87                         |

TABLE IV
COMPARISON FROM F-MEASURE

|             | F-mesure  |                                   |
|-------------|-----------|-----------------------------------|
|             | E-Rec-Sys | (Amdouni and Ben Abdesslem, 2010) |
| Name        | 0.86      | 0.85                              |
| Gender      | 0.89      | 0.88                              |
| Nationality | 0.98      | 0.91                              |
| Date        | 0.93      | 0.9                               |
| Adress      | 0.83      | 0.83                              |
| Phone       | 0.99      | 0.97                              |
| Email       | 0.87      | 0.88                              |
| Institute   | 0.82      | 0.79                              |
| Language    | 0.97      | 0.84                              |

However, according to Table (Table III), the F-measure of E-Rec-Sys is better then Amdouni and Ben Abdessalem work in annotating the concepts of : Name, Gender, nationality, date, tel, establishment and language. Nevertheless, Amdouni and Ben Abdessalem work is better in annotating: Email, competence, and training with values very close to ours.

We wanted to test whether corpus size influences the response time of annotation. We measured the response time of our system during the annotation process of resumes for 1, 10, 50, 100, 150, …, and total 300 resumes.
The result is presented in the table above (TableV).

TABLE V
RESPONSE TIME

| Resume number | Response time |
|---|---|
| 1 | 10s |
| 10 | 1mn 15s |
| 50 | 3mn 10s |
| 100 | 4mn 45s |
| 150 | 6mn 35s |
| 200 | 8mn 45s |
| 300 | 11mn 14s |

These results show that the response time during the annotation process is not strongly influenced by the corpus size, for 10 resume, for example, the response time is 1mn 15s, which is lower than 10 resume* 10s.

## 5. Conclusion

Employers often collect a huge quantity of resumes for an open post. In order to decrease management costs employers have begun to choose online resumes rather then classic ones. The resumes are used for the automated selection of candidates satisfying the requirements. For this purpose annotations are useful for facilitating the selection process by matching the requirements with the resumes. We claim that using Semantic Web technologies (ontology) is helpful for the resume annotations.

In this paper we presented a scenario for supporting recruitment processes. We suggested an annotation approach of resumes based on ontology and we described our prototypical implementation.

We will focus our future work on the refinement of the prototype, and integrating a component for the semantic matching between demand (resume) and supply (requirement).